\documentclass[fleqn,twoside]{article}
\usepackage{espcrc2}

\usepackage{graphicx}
\usepackage[figuresright]{rotating}

\newcommand{\AmS}{{\protect\the\textfont2
  A\kern-.1667em\lower.5ex\hbox{M}\kern-.125emS}}

\title{Extreme Physics via X-rays from Black Holes and
`Neutron' Stars}

\author{Martin Elvis\address[CfA]{Harvard-Smithsonian Center for
 Astrophysics, 60 Garden St., Cambridge MA 02138 USA}
\thanks{elvis@cfa.harvard.edu}
}

\begin{document}

\title{Extreme Physics via X-rays from Black Holes and
`Neutron' Stars}
\date{2004/2/1}

\begin{abstract}
A combination of microchannel plate optics and a 32$\times$32
pixel microcalorimeter would allow the successor to the
{\em Rossi XTE} to explore new domains of spectroscopic timing in
a MIDEX class mission. With $\sim$10 times the area and $\sim$100
times the spectral resolution of the PCA (and 10 times that of
silicon detectors) such a mission would be able to explore
redshifts and plasma conditions in weak line features over a wide
range of celestial sources. This would allow several tests of
basic physics, both QED/QCD and GR.
\vspace{-0.1in}
\end{abstract}

\maketitle

\section{AN EXTREME PHYSICS MISSION}

Question~6 in the Turner report\cite{Turner01} {\em Connecting Quarks
to the Cosmos} asks {\bf ``What are the limits of physical law?''}.
The report goes on to point to black holes and neutron stars as the
places that could best test these limits. Both types of compact object
are seen primarily through X-ray radiation emitted close to their
surface or event horizon, derived from gas accreted onto them from a
closely orbiting companion star.  In principle these X-rays could
probe the physics near these exotic objects. Here we show how to
tackle Question~6 head on, using a novel combination of technologies
to build an X-ray detection system for the signals from these cosmic
laboratories.

By combining high spectral resolution with high time resolution this
mission allows the study of weak emission and absorption lines, and
potentially polarization, in celestial X-ray sources. In this way we
can probe matter under conditions more extreme than could be produced
in any terrestrial laboratory for strong gravitational or magnetic
field. Spectral lines are especially good ``clocks'' for GR, and are
sensitively affected by the ultra-strong ($\sim$10$^{15}$g) magnetic
fields in magnetars (Table~1).  For example in QED compression of the
wavefunction perpendicular to the B-field in magnetars will move the
Lyman series up to X-ray wavelengths.

Space missions that address fundamental physics tend to be single
experiments. This mission would, in contrast, operate as a fundamental
physics facility with multiple experiments, like Fermilab or CERN, but
with X-ray sources, such as the black hole Cygnus~X-1 or the magnetar
SGR 1806-20, taking the place of the accelerator.
As {\bf a physics mission} this satellite would only produce X-ray
binary astrophysics incidentally, although we expect that a great deal
of X-ray binary astro-physics will be a by-product of the primary
mission.
Physics topics that could be addressed span both pillars of 20$^{th}$
century physics, Quantum Electrodynamics and General Relativity.  In
Table~1 we begin an enumeration of the extreme physics that could be
addressed.

\begin{table}[t]
\caption{Extreme Physics with Black Holes, Neutron Stars and Magentars}
\label{tab:a}
\begin{tabular}{ll}
\hline
Physics&Sample Method\\
\hline
QED in Strong&Magnetars\cite{Harding03}\cite{Miller01}\\
magnetic fields&\\
\hline
QCD Strange& Equation of State\\
(Quark?) Matter&of Neutron Stars\cite{Henning02}\\
\hline
GR frame & black hole QPOs, lines\\
dragging& \cite{Stella98}\\
\hline
GR metric in & Fe-K lines, and \\
strong gravity &low energy `satellite' lines, \\
&in X-ray binaries, AGNs\cite{Miller02}\\
\hline
\end{tabular}
\end{table}

The {\em Rossi X-ray Timing Explorer, RXTE} has had a tremendous
impact on our knowledge of the properties of these objects which have
the most extreme physics in the Universe.  The asrophysics of these
objects - their gas flows and radiation emission mechanisms - are now
understood at a level of detail unimagined 10 years ago.  Instead of
being mysterious objects of investigation, X-ray binaries are becoming
tools for the investigation of fundamental physics near neutron stars
and black holes.
To realize this potential requires a qualitative leap beyond {\em
RXTE}.  Precise spectral measurements beyond the $\sim$10\% resolution
of RXTE above 2~keV, or XMM-Newton at $\sim$1~keV are essential, yet
with many times the collecting area of {\em Chandra}, to collect
sufficient photons.

The value of spectral resolution is shown by the goal of measuring the
radius of a neutron star and so determining its equation of
state. This will tell us if these are truly `neutron' stars or
something more exotic: strange stars. The gravitational redshift of an
atomic line feature from a neutron star surface measures the radius of
the star, and hence the constrains the equation of state. Such a line
has been seen, but to limited accuracy. A factor 10 or more
improvement in spectral resolution would let us test models.
The need for collecting area is illustrated, for example, by
Quasi-Periodic Oscillations (QPOs). Since QPOs in X-ray binaries could
be the key to exploring General Relativity (GR) in the strong gravity
limit, this matters. The shape of QPOs is ambiguous, since we know
them only through their power spectra, and co-adding is not possible
because they are only {\em quasi}-periodic oscillations. Some
5~m$^2$-10~m$^2$ is needed to see individual QPOs.

\section{THE MISSION CONCEPT}

We propose a MIDEX class mission concept to achieve these goals.
Microchannel plate (MCP) optics have a factor 100 lower weight to
effective area ratio compared with ASCA-like foil optics, allowing the
needed 5-10sq.meter mirror.  At the single focus of the MCP optics
would sit a microcalorimeter detector with 10-50 times the spectral
resolution of the RXTE or of silicon detectors.  Microcalorimeter are
infeasible for non-imaging or multi-focus configurations.  The mission
will spend long intervals pointed at a single target. High orbits are
desirable for uninterrupted viewing of sources for Fourier analysis.
Table~2 shows how the mission fits within the weight envelope for a
Delta~2 launcher. [The Con-X XMS calorimeter has a mass budget of
123~kg, including 90~kg for the cryo system \cite{Gadwal99}.]
Compared with Constellation-X, this mission has major weaknesses:
poorer angular resolution is {\em a requirement} (for the calorimeter
readout, see \S 2.2), no diffraction gratings, no high energy
telescope. The power of the mission comes from concentrating on a niche
application from an astronomy point of view, but a major field, when
considered as a physics experiment.

The low background that comes from arcminute imaging would allow the
mission to explore a wide range of celestial sources including some of
the bright ``Ultra-Luminous X-ray Sources'' (ULXs) in external
galaxies, which may be intermediate mass ($\sim$10$^3$M$_{\odot}$)
black holes, and bright AGN hosting $\sim$10$^6$-10$^9$M$_{\odot}$
black holes.

\begin{table}[t]
\caption{Strawman Mission Mass Budget}
\label{tab:b}
\begin{tabular}{lr}
\hline
{\em Science Payload}&{\em Mass}\\
\hline
10~m$^2$ Microchannel Plate Optics&37~kg\\
Mirror support structure&37~kg\\
30~m Optical Bench&20~kg\\
Calorimeter + cryo system&123~kg\\
\hline
{\em Spacecraft}&200~kg\\
\hline
20\% Reserve&83~kg\\
\hline
TOTAL&500~kg\\
\hline
\end{tabular}
\end{table}

\subsection{Microchannel Plate Optics}

The key technology is MCP optics.  With their area:mass advantage a
10~m$^2$\cite{Fraser97} the MCP mirror would weigh only 37~kg (plus an
equal mass support structure).  MCP optics have high aperture
utilization ($>$75\%), and the dense packing of MCP optics close to
the optical axis gives significant response above 10keV, which could
be increased by using a longer optical bench.  Arcminute resolution
has been demonstrated at 8~keV in a radially packed geometry with a
circular PSF\cite{Bavdaz02}.  A 25~meter focal length with the same
focal ratio as {\em Chandra} would have a 3~m dia mirror and so a
geometrical area of 7.1~m$^2$, or an open area of 5.65~m$^2$ for a
plausible 80\% aperture utilization. Reflection efficiency would
reduce this to $\sim$3.75~m$^2$ up to 2~keV. Increasing the focal
length to 40~m and mirror diameter to 5~m would give 8~m$^2$ {\em
effective} area in the few keV range.  The plate-like geometry of
microchannel plate optics allows them to be folded for launch in a
compact configuration, and then deployed with simple mechanisms
\cite{Bavdaz02}.
The $\sim$30~meter focal length is readily achieved arcminute imaging
level with flight tested extendable structures. The Able Engineering
Company (AEC\cite{AEC}) has supplied lightweight (Table~2)
continuous-longeron coilable booms of similar lengths to several
missions (UARS, GGC WIND, GGS POLAR, Cassini, Lunar Prospector, IMAGE).

\subsection{Rapid Readout Microcalorimeter}

A 10~m$^2$ mirror looking at the brightest X-ray sources needs a rapid
readout detector. Typical neutron star and black hole X-ray binaries
yield 10$^4$ct~s$^{-1}$.  Microcalorimeters have a reasonable maximum
readout rate {\em per pixel} \cite{Stahle99}\cite{Silver00} of
1000~Hz. To avoid pile-up and the resultant dead time and
spectroscopic/timing degradation effects, the detectors need to have a
readout rate about 10 times the count rate.  This is achieved by
spreading the signal over $>$100 pixels. Such arrays are now
achievable.  The XRS on ASTRO-E2 (launch February 2005) has a
6$\times$6 array. Con-X is aiming for a 32$\times$32
array\cite{Kelley03}. The plate scale for a 30~meter focal length
optic is 150$\mu$m/arcsecond, so a 1~arcminute PSF would spread over
9~mm.  An 18~mm dia. 32$\times$32 array would have a pixel size of
$\sim$600~micron. The XRS is $\sim$4~mm dia and so only modest changes
in size and heat capacity are required.

\subsection{Polarimeter}

X-ray polarimetry can provide a whole new diagnostic for the physics
in high magnetic and gravitational fields \cite{Heyl03},
\cite{Blandford01}.  But polarization measurements are photon
hungry. To measure a typical polarization of 1\% to $\pm$10\% needs
10$^6$ photons (while a 10\% total flux measurement needs only 100
photons) and so has never been achieved. A 10~m$^2$ mirror is just
what is needed. The large field of view of MCP optics, compact dewar
designs and closed cycle refrigerators may allow an X-ray polarimeter
\cite{Costa01} to be included.

\section{EXTREME PHYSICS EXPLORER: A FITTING SUCCESSOR TO {\em Rossi
XTE}} 

This concept is of an {\bf extreme physics mission}, achieved through
opening up orders-of-magnitude of new timing/spectroscopy/polarimetry
territory.  Yet the mission fits within the MIDEX envelope, and so can
plausibly be brought to fruition on a modest timescale.  The next step
is to have community meetings focussed on the physics to be done, and
on the mission technologies that can make it happen. Such meetings
needs to involve the high energy physics community and the high energy
astrophysics community working together.




\end{document}